\documentclass[12pt,preprint]{aastex}

\usepackage{amsmath}
\usepackage{amssymb}

\newcommand{\be}{\begin{equation}}
\newcommand{\ee}{\end{equation}}
\newcommand{\bea}{\begin{eqnarray}}
\newcommand{\eea}{\end{eqnarray}}

\shorttitle{The small-scale Universe}
\shortauthors{Widrow, Elahi, \& Thacker}

\begin{document}

\title{Power spectrum for the small-scale Universe}
\author{Lawrence M. Widrow\altaffilmark{1}, 
Pascal J. Elahi\altaffilmark{2}
\affil{Department of Physics, Engineering Physics, and
Astronomy, Queen's University,
Kingston, ON, K7L 3N6, Canada}}
\author{Robert J. Thacker\altaffilmark{3}, Mark 
Richardson\altaffilmark{4}}
\affil{Department of Astronomy and Physics, Saint Mary's University,
Halifax, NS, B3H 3C3, Canada}
\and
\author{Evan Scannapieco\altaffilmark{5}}
\affil{School of Earth and Space Exploration, Arizona State University,
PO Box 871404, Tempe, AZ, 85287-1404}

\altaffiltext{1}{widrow@astro.queensu.ca}
\altaffiltext{2}{pelahi@astro.queensu.ca}
\altaffiltext{3}{thacker@ap.stmarys.ca}
\altaffiltext{4}{mrichard@ap.stmarys.ca}
\altaffiltext{5}{evan.scannapieco@asu.edu}

\begin{abstract}

  The first objects to arise in a cold dark matter universe present a
  daunting challenge for models of structure formation.  In the ultra
  small-scale limit, CDM structures form nearly simultaneously across
  a wide range of scales.  Hierarchical clustering no longer provides
  a guiding principle for theoretical analyses and the computation
  time required to carry out credible simulations becomes prohibitively high.
  To gain insight into this problem, we perform high-resolution
  ($N=720^3 - 1584^3$) simulations of an Einstein-de Sitter cosmology
  where the initial power spectrum is $P(k) \propto k^n,$ with
  $-2.5\le n \le -1$.  Self-similar scaling is established for $n=-1$
  and $n=-2$ more convincingly than in previous, lower-resolution
  simulations and for the first time, self-similar scaling is
  established for an $n=-2.25$ simulation.  However, finite box-size
  effects induce departures from self-similar scaling in our $n=-2.5$
  simulation.  We compare our results with the predictions for the
  power spectrum from (one-loop) perturbation theory and demonstrate
  that the renormalization group approach suggested by
  \citet{mcdonald} improves perturbation theory's ability to predict
  the power spectrum in the quasilinear regime.  In the nonlinear
  regime, our power spectra differ significantly from the widely used
  fitting formulae of \citet{pd96} and \citet{smith} and a new fitting
  formula is presented.  Implications of our results for the stable
  clustering hypothesis vs.~halo model debate are discussed.  Our
  power spectra are inconsistent with predictions of the stable
  clustering hypothesis in the high-$k$ limit and lend credence to the
  halo model.  Nevertheless, the fitting formula advocated in this
  paper is purely empirical and not derived from a specific
  formulation of the halo model.

\end{abstract}



\section{Introduction}

In the standard cosmological paradigm, present-day structures arise
from small-amplitude density perturbations which have their origin in
the very early Universe.  These primordial perturbations are presumed
to form a Gaussian random field whose statistical properties are
described entirely by the power spectrum, $P(k)$.  While higher order
statistics are required to describe the density field once
nonlinearities develop, the power spectrum remains central to our
understanding of structure formation.

During the radiation-dominated phase of the standard cold dark matter
(CDM) scenario, the power spectrum evolves from its primordial,
approximately power-law form, $P(k)\propto k$ to one in which its
logarithmic slope, $n_{\rm eff}\equiv d\ln P(k)/d\ln k$, decreases
from $n_{\rm eff}\simeq 1$ on large scales to $n_{\rm eff}\simeq -3$
on small scales\footnote{More precisely, the logarithmic slope decreases
  from $n_{\rm eff}=n_p$ to $n_{\rm eff}=n_p~-~4~+$ {\it logarithmic
    corrections} where $n_p$ is the spectral index of the primordial
power spectrum.  An analysis of the three-year WMAP data by
  \citet{wmap} indicates that $n_p = 0.958\pm 0.016$.  However, for
  the sake of argument, we will set $n_p=1$ since the difference is
  not relevant for our discussion.}.  At the start of the
matter-dominated phase, which signals the beginning of structure
formation, the dimensionless power spectrum, $\Delta^2(k)\propto
k^3P(k)$, decreases monotonically with scale.  The implication is that
structure forms from the bottom up.  Hierarchical clustering, as this
process has come to be known, is the central idea in our understanding
of structure formation.  Hierarchical clustering also explains why
cosmological N-body simulations are able to provide a reasonable
facsimile of true cosmological evolution; with enough dynamic range,
simulations are able to follow the development of virialized, highly
nonlinear systems on small scales while properly modeling the
large-scale tidal fields that shape them.  However, the dynamic-range
requirement becomes increasingly difficult to achieve as $n_{\rm
  eff}\to -3$ since, in this limit, $\Delta^2$ becomes independent of
$k$ and structures collapse nearly simultaneously across a wide range
of scales.  Put another way, as $n_{\rm eff}\to -3$, the infrared
divergence of the power spectrum becomes increasingly problematic for
numerical (as well as theoretical) studies.

Interest in the small-scale limit of the CDM hierarchy was prompted by
the realization that dark matter halos have a wealth of substructure
\citep{moore99a, klypin} and that this substructure may have important
implications for both direct and indirect dark matter detection
experiments (See, for example, \citet{stiff, diemand, kuhlen} and
\citet{kamionkowski}).  High-resolution simulations suggest that the
subhalo mass function extends down to the dark matter free-streaming
scale with approximately constant mass in substructure per logarithmic
mass interval.  These simulations probe structures which form from an
initial power spectrum with $-3< n_{\rm eff}<-2$.  For example, in
the simulation of the first CDM objects by \citet{diemand}, $n_{\rm
  eff}\simeq -2.8$.  With such extreme spectra, care must be taken in
order to insure that the results are not corrupted by finite-volume
effects.  This issue, as it relates to the halo and subhalo mass
functions, is discussed in \citet{power} and \citet{bagla} as well as
in the companion to this paper, \citet{elahi}.

In this paper, we provide insight into the small-scale limit of CDM by
focusing on scale-free cosmologies, that is Einstein-de Sitter
cosmologies where the initial power spectrum is a power-law function
of $k$, $P(k)\propto k^n$.  The guiding principle for understanding
structure formation in these models is self-similar scaling which
implies that the functional form of the dimensionless power spectrum
is time-independent, up to a rescaling of the wavenumber $k$ (see
below).  Self-similar scaling provides a diagnostic test of whether a
simulation has sufficient dynamic range \citep{jain98}.  Contact with
the standard $\Lambda$CDM cosmology is made by treating parameter $n$
as a proxy for scale: $n\simeq -1.8$ corresponds to cluster scales and
$n\simeq -2.2$ to galactic scales.  The limit $n\to -3$ corresponds
to the bottom of the CDM hierarchy.

We perform N-body simulations with $n=-1,-2,-2.25,$ and $-2.5$, and
compare our results directly with theoretical models.  The computation
costs to conduct credible simulations with $n < -2.5$ are
prohibitively high (see below) and even our $n= -2.5$ results must be
considered suspect because of finite box-size effects.  We compare our
results for the power spectra in the quasilinear regime with
predictions from one-loop perturbation theory and demonstrate that for
$n=-2$ and $-2.25$, the agreement is vastly improved if one implements
the renormalization-group approach suggested by \citet{mcdonald}.

To predict the full non-linear power spectrum, one must resort to
semi-analytic models such as the ones described in \citet{hklm} and
\citet{pd96}.  These models are based on the stable-clustering
hypothesis \citep{pjep74,dp77} which holds that gravitationally-bound
systems decouple from the rest of the Universe once they collapse.  An
alternative, known as the `halo model' \citep{mafry, ps00, seljak},
allows for the continual accretion of mass onto existing haloes.  The
density field is treated as a distribution of mass concentrations,
each characterized by a density profile.  The power spectrum then
involves the convolution of this density profile with the halo mass
function.  Simulations by \citet{smith} of structure formation in a
number of scale-free cosmologies demonstrate a clear departure from
the stable clustering hypothesis and appear to support the halo model,
at least qualitatively.  

\citet{pd96} and \citet{smith} provide fitting formulae for nonlinear
power spectra.  Formally, these formulae apply to all initial power
spectra with $n>-3$ though they are calibrated using simulations with
$n\ge -2$.  One goal of this paper is to provide an alternative fitting 
formula which applies when $n\le -2$.

The overall layout of this paper is as follows:  In \S 2, we present
background material including a discussion of self-similar scaling,
perturbation theory, and semi-analytic models.  We describe our
simulations in \S 3 and our results in \S 4.  We also provide a
new and improved fitting formula for the nonlinear power spectra.  In
\S 5, we discuss the implications of our results for the halo model
and stable clustering hypothesis.  We conclude, in \S 6, with a
summary and a discussion of directions for future investigations.

\section{Preliminaries}\label{prelim}

\subsection{Statistics of the Density field}

In keeping with standard definitions ({\em e.g.} Peacock 1999), we
express real-space density perturbations as deviations from the mean
background density, $\rho_{\rm bg}(t)$, and then construct the $k$-space
representation as follows:
\begin{align}
\delta\left ({\bf x},\,t\right ) &= \frac{\rho\left ({\bf x},\,t\right )
- \rho_{\rm bg}(t)}{\rho_{\rm bg}(t)}\\
& = \int \frac{d^3k}{\left (2\pi\right )^3}\delta\left ({\bf k},t\right )
e^{i{\bf k}\cdot{\bf x}}~.
\end{align}
The power spectrum, $P(k)$, (strictly speaking a spectral density) is
defined by the relation,
\begin{equation}
\left (2\pi\right )^3\delta_D\left ({\bf k}-{\bf k}'\right )
P\left ({\bf k'}\right ) = \langle \delta\left ({\bf k}\right )
\delta\left ({\bf k'}\right )\rangle,
\end{equation}
where $\langle\cdots\rangle$ denotes an ensemble average, $\delta_D$
is the Dirac delta function, and where the $t$-dependence is dropped
for notational simplicity.  Note that both $\delta \left ({\bf
    k},\,t\right )$ and $P\left ({\bf k},\,t\right )$ have units of
volume.  So long as the density field is statistically isotropic,
$k=|{\bf k}|$ encapsulates the wavenumber dependence.  A useful
quantity is the dimensionless power spectrum, $\Delta^2(k)\equiv
k^3P(k)/2\pi^2$, which measures the power per logarithmic wavenumber
bin.  Given these definitions, the wavenumber at which $\Delta^2\sim
1$ corresponds to the dividing line between the linear and nonlinear
scales.

\subsection{Initial Conditions}

Initial conditions for our N-body simulations are specified at an
early epoch when the density field is accurately described by linear
perturbation theory.  We assume that prior to this epoch, the power
spectrum is a power-law function of $k$ between appropriately chosen
high and low wavenumber cutoffs.  That is,
\begin{equation}
P\left (k,a\right ) = P_L\left (k,\,a\right )f_{UV}\left (k,k_c\right )
f_{IR}\left (k,k_B\right )
\end{equation}
where
\begin{equation}
\label{eq:plinear}
P_{L}(k,a) = Aa^2k^n~,
\end{equation}
$a$ is the scale factor, and $A$ is a normalization constant.  (Here
and throughout, we assume an Einstein-de Sitter cosmology.  For other
cosmological models, $a$ is replaced by the linear growth factor,
$D(a)$.)  The functions $f_{UV}$ and $f_{IR}$ truncate the power
spectrum above $k_c$ and below $k_B$, respectively.  We assume $k_B =
2\pi/B$ and $k_c = k_{Ny} = \pi N^{1/3}/B$ where $B$ is the size of
the simulation ``box'' and $k_{Ny}$ is the Nyquist wavenumber of the
initial particle distribution.  We also follow \citet{kudlicki} in
using the truncation functions
\begin{equation}
f_{UV}(k,k_c) = e^{-(k/0.8 k_c)^{16}} \;\;\;\;\;\; 
f_{IR}(k,k_B)=\Theta(k-k_B).
\end{equation}
where $\Theta(x)$ is the Heaviside function.  A similar initial set-up
is used in our renormalization group calculations (see section
\ref{qlregime}).  Our assumptions imply that at this early epoch
$\Delta^2\left (k_c\right )\ll 1$, so that all modes within the
simulation or computation box are initially in the linear regime.

Utilizing these definitions, we describe the nonlinear scale 
$L_{NL}=2\pi/k_{NL}$ by the condition
$\Delta_L^2(k=k_{NL},\,a) = 1$, {\em i.e.}, $k_{NL} = \left (4\pi A
  a^2\right )^{-1/\left (n+3\right )}$.  Since $L_{NL}$ is the only
preferred length scale, the power spectrum should evolve according to
the self-similar scaling ansatz
\begin{equation}\label{eq:selfsim}
\Delta^2\left (k,a\right ) = \widehat{{\Delta}^2}\left (k/k_{NL}\right )
\end{equation}
(see \citet{jain96, jain98} for a discussion and review of the
literature).  Note that since $\Delta_L^2 = \left (k/k_{NL}\right
)^{n+3}$, we can also write $\widehat{\Delta^2}$ as a function of
$\Delta_L^2$.

\subsection{Perturbation Theory}

The evolution of the power spectrum from the linear regime to the
mildly nonlinear or quasilinear regime can be estimated via
perturbation theory (PT, see, for example, \citet{bernardeau}). The
starting point is an expansion for the density perturbation field of
the form:
\begin{equation}
  \delta\left ({\bf k},t\right ) = \sum_{n=1}^\infty = a^n(t)\,
  \delta_n\left ({\bf k}\right )
\end{equation}
where $a$ is again the scale factor, $\delta_1$ characterizes linear
density fluctuations, and $\delta_n$ denote terms of order
$\left(\delta_1\right )^n$.  So long as the density field is
statistically isotropic, the perturbative expansion can be written in
terms of the power spectrum:
\begin{equation}
P_{PT}(k,a) = P_{L}(k,a) + P^{(1)}(k,a) + \dots
\end{equation}
where $P^{(1)} = O\left (k^3P_L^2\right )$.  In practice, the series
is rarely carried beyond second order in $P_L$.  The calculations are
aided by a diagrammatic scheme in which the higher-order terms are
described as ``loop corrections'' to the ``tree-level'' term, $P_L$
\citep{sco96}.  The one-loop correction, $P^{(1)}$, comprises two
distinct terms or diagrams, $P_{13}$ and $P_{22}$.  These terms
involve integrals over the linear power spectrum, $P_L$.  Explicit
expressions can be found in \citet{sco96}, \citet{mcdonald} and
elsewhere.  For scale-free models, they combine to give
\begin{equation}\label{eq:deltaPT}
  \Delta^2_{PT}\left (k,\,a\right ) = \Delta^2_L(k,a)
  \left ( 1 + \lambda(n)\Delta^2_L(k,a)\right ) + 
  O\left (\left (\Delta_L^2\right )^3\right )
\end{equation}
where $\lambda(n)$, which can be calculated analytically, is positive
for $n>-1.4$ and negative for $n<-1.4$ \citep{sco96, bernardeau}.
Hence, $n\simeq -1.4$ represents a ``critical index'' where nonlinear
corrections are vanishingly small \citep{rs97}.

\subsection{Renormalization Group Approach}\label{RGPT}

PT breaks down when the loop corrections, which are formally
divergent, become comparable to the tree-level terms.  The
renormalization group (RG) scheme proposed by \citet{mcdonald}
alleviates this problem essentially by updating the perturbative
expansion as the system evolves.  RG removes the divergences in the PT
expansion, leaving behind a well-behaved, renormalized power spectrum.
Operationally, one begins with an initial power spectrum and takes a
small step forward in time using one-loop perturbation theory.  The
new power spectrum is used as the initial condition for the next
timestep.  This procedure is accomplished by solving the
integro-differential equation
\begin{equation}\label{eq:rgequation}
\frac{d\tilde{P}}{d\left (a^2\right )} = \tilde{P}_{13} + \tilde{P}_{22}
\end{equation}
with the initial conditions, $\tilde{P} = P_L/a^2$.  Here,
$\tilde{P}_{13}=P_{13}/a^2$ and $\tilde{P}_{22}=P_{22}/a^2$ with the
proviso that $\tilde{P}$ rather than $P_L$ is used in evaluating
$\tilde{P}_{13}$ and $\tilde{P}_{22}$ (see \citet{mcdonald}).  Details
on our own scheme for solving this equation are given in Section
\ref{qlregime}.

As discussed in \citet{mcdonald}, the method has a number of limitations.
In particular, Eq.\,\ref{eq:rgequation} ignores higher-order terms
(two-loop and beyond) in $P_L$.  Moreover, decaying mode solutions
are not included.  Thus, our RG results should be interpreted with
a degree of caution.

The approach described here is an example of how RG techniques can be
used to remove secular divergences in differential equations.
\citet{crocce06} and \citet{crocce08} outline an alternative method to
study the nonlinear evolution of large-scale structure which also
employs RG techniques.  Their formalism is conveniently represented in
terms of Feynmann diagrams and is closer in spirit to RG applications
in high energy and statistical physics.

\subsection{Nonlinear Regime --- Stable Clustering vs. Halo Model}

N-body simulations provide the most direct means to determine the
nonlinear power spectrum.  The so-called HKLM procedure provides an
avenue by which one can begin to understand the simulation results
from a theoretical basis \citep{hklm, pd96}.  The approach yields
fitting formulae for the nonlinear power spectrum (or two-point
correlation function, $\xi(r)$) in different cosmological models.  The
key assumption is the existence of a one-to-one relation between the
nonlinear power spectrum at wavenumber $k$ and the linear power
spectrum at an earlier epoch and smaller wavenumber, $k_L$.  The
starting point is the mapping
\begin{equation}
\label{eq:kmapping}
k_L = \left (1 + \Delta^2(k,a)\right )^{-1/3}k,
\end{equation}
which is the analogue of the real-space relation,
\begin{equation}
\label{eq:rmapping}
r_L = \left (1 + \bar{\xi}(r)\right )^{1/3}r~,
\end{equation}
that results from associating the volume averaged correlation
function, $\bar{\xi}$, with the local over-density.  The nonlinear
power spectrum may then be written as a function of the linear power
spectrum at the earlier epoch:
\begin{equation}
\label{eq:HKLM}
\Delta^2(k,a) = f\left(\Delta^2_L(k_L)\right ),
\end{equation}
where $f$ is an appropriately chosen fitting formula and, by
assumption, $\Delta^2(k_L)\ll 1$.  Note that for scale-free models,
Eq. \ref{eq:kmapping} together with Eq. \ref{eq:selfsim} implies
Eq. \ref{eq:HKLM}.  In the linear regime $f(x)$ must be the identity
function, {\em i.e.}, $f(x) = x$, and the power spectrum grows with
the scale factor as $\Delta_L^2\propto a^2$. In the nonlinear regime,
\citet{hklm} and \citet{pd96} appeal to the ``stable clustering
hypothesis'', which holds that highly nonlinear structures decouple
from the expansion.  Under this assumption, $\Delta^2\propto a^3$ and
the asymptotic behaviour of $f(x)$ must be given by $f\propto
x^{3/2}$.

To estimate $f(x)$, \citet{pd96} advocate the fitting formula
\begin{equation}\label{eq:pd}
f(x) = x\left \{\frac{1 + B\beta x + \left (Ax\right )^{\gamma\beta}}
{1 + \left (Ax\right )^{\gamma}/\left (Vx^{1/2}\right )^\beta}
\right \}^{1/\beta}
\end{equation}
where the parameters $\gamma$, $\beta$, $V,~A,$ and $B$ are determined
by fitting Eq.\,\ref{eq:pd} to power spectra measured in
simulations. The parameters are understood as follows: $B$ determines
a second-order departure from linear growth, $A$ and $\gamma$ control
the behaviour in the quasilinear regime, $V$ controls the amplitude of
the asymptote and $\beta$ shapes the transition between the two
regimes. Best-fit parameters are expressed as functions of $1 + n/3$,
for example, $\gamma=3.310\left (1+n/3\right )^{-0.244}$.  The
expressions for the other parameters similarly diverge as $n\to -3$
though one must bear in mind that they are based on results from
simulations with $n\ge -2$.

The assumption of stable clustering has been challenged by various
groups \citep{mafry,ps00,seljak,smith} on the basis that dark matter
haloes continually accrete matter and never fully decouple from the
rest of the Universe.  An alternative approach is provided by the halo
model in which the density field is given as a distribution of mass
concentrations (haloes) which evolve and have their own internal
structure.  The two-point correlation function comprises a one-halo
term, which is associated with the correlation of mass within a single
halo, and a two-halo term, which is associated with the correlation
between different haloes \citep{mafry,ps00,seljak,smith}.  Since the
power spectrum is the Fourier transform of the two-point correlation
function, the associated components of the power spectrum, $P_{1h}$
and $P_{2h}$, involve integrals over the halo mass function and
Fourier-transformed halo density profile.  We return to this point in
Section 5.

Motivated in part by the separation of components used in the 
halo model, \citet{smith} construct a fitting formula in the form
\begin{equation}
\Delta^2\left (k,\,a\right ) = \Delta^2_Q\left (k,\,a\right )
+ \Delta^2_{NL}\left (k,\,a\right )
\end{equation}
for the power spectra measured in their simulations.  By construction,
$\Delta_Q^2$ dominates the power spectrum in the quasilinear regime
and is meant to account for halo-halo correlations while
$\Delta^2_{NL}$ dominates the power spectrum in the nonlinear regime
and is meant to account for single halo correlations.  However, the
model is purely empirical and not calculated directly from the halo
model.  It is also worth noting that their formula has eight free
parameters, three more than that of \citet{pd96}.

\section{Simulations}

N-body simulations are carried out with scale-free initial conditions
and $n=-1$, $-2$, $-2.25$, $-2.5$ using the parallel tree-PM code
GADGET-2 \citep{gadget2}.  Initial conditions are generated on a
regular grid using a second order-Lagrangian Perturbation Theory
(2LPT) code \citep{crocce,rjt06}.  The primary benefit of 2LPT is to
reduce the impact of spurious transient modes which arise from the
truncation of the perturbative expansion.  Since these modes decay,
their impact is to delay the time in the simulation at which credible
statistics can be calculated \citep{rs98,crocce}.  The
  simulations are run with a softening length of $1/30$ the initial
  interparticle spacing.  GADGET-2 parameters such as the opening
  angle used in constructing the particle tree and maximum time step
  criterion were set to their default value.

Table \ref{tab:sims} summarizes key features of the simulations used
in this study such as the number of simulation particles and the
epochs at which the power spectra are measured.  The latter are
  expressed in terms of the ratio $a/a_*$ where $a_*$ is defined as
  the scale factor at the epoch when the mode on the scale of the box
  is equal to the nonlinear scale, that is, $a/a_* = \left
    (k_B/k_{NL}\right )^{\left (n+3\right )/2}$.  With this
  definition, $\left (a/a_*\right )^2 = \Delta^2\left (k_B\right )$.

\begin{figure}
\epsscale{1.0}
\plotone{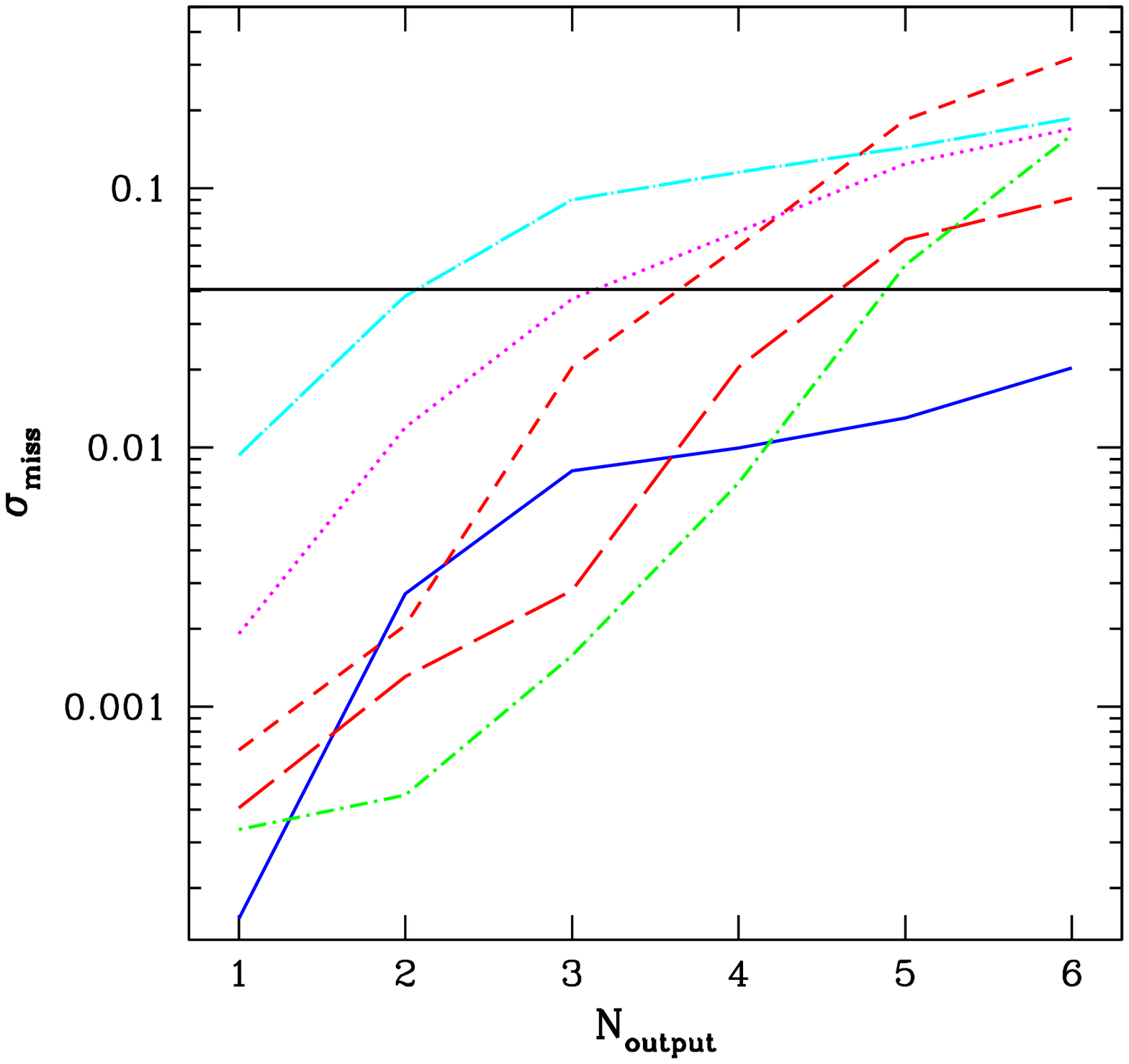}
\caption{Missing variance, $\sigma_{\rm miss}$, as given in
  Eq.\ref{eq:sigmamiss}, for the six simulations described in the text
with line types/colours as follows:
solid/blue --- $n=-1$; dotted/red --- $n=-2$, $N=32^3$; short-dashed/red ---
$n=-2$, $N=256^3$; long-dashed/red --- $n=-2$, $N=1024^3$; 
short-dashed-dot/green --- $n=-2.25$; long-dashed-dot/cyan ---
$n=-2.5$.  The output number, $N_{\rm output}$ corresponds to
  the values listed in Table 1.  For $n=-2.5$, we show $\sigma_{\rm
    miss}$ for the last six outputs.  The horizontal black curve corresponds
to the criterion adopted by \citet{smith}.}
\label{fig:sigmamiss}
\end{figure}

  As $n$ approaches $-3$, the absence of modes beyond the box scale
  induces an error in the nonlinear power spectrum.  \citet{smith}
  adopt the {\it ad hoc} criterion that the missing variance,
  $\sigma_{\rm miss}$, associated with these modes satisfies the
  condition
\begin{equation}\label{eq:sigmamiss}
\sigma_{\rm miss}\le 0.04
\end{equation}
where, to a good approximation, $\sigma_{\rm miss} =
\Delta_L^2\left (k_B\right )G(3+n)$ with $G(y) = (1 - 0.31y +
0.015y^2 + 0.00133y^3)/y$.  Figure \ref{fig:sigmamiss} shows
$\sigma_{\rm miss}$ for the outputs of our six simulations.  We see
that the final two outputs in our high-resolution $n=-2$ and $n=-2.25$
simulations and all but the first few outputs in our $n=-2.5$
simulation fail to meet the \citet{smith} criterion.  We return to
this point below.

Power spectra are calculated using a cubic mesh with side length $L$
(Table 1).  We set $L=2N_m$ so long as $N_m$ is a power of 2. Here,
$N_m=N^{1/3}$ is the side length of the initial grid of simulation
particles.  When $N_m$ is not a power of 2 we set $L$ equal to the
first power of 2 larger than $N_m$.  The power spectra are calculated
using piecewise quadratic spline interpolation \citep{he81} and
adjusted to account for the strong filtering of this mass-assignment
scheme.  No correction is made for shot noise.

\begin{deluxetable}{ccccc}
\tablecaption{Summary of Simulations \textbf}
\tablewidth{0pt}
\tablehead{
\multicolumn{1}{c}{$n$} & \multicolumn{1}{c}{$N$} & $L$ &
\multicolumn{1}{c}{initial scale factor $(a_i/a_*)$} &
\multicolumn{1}{c}{output scale factors $(a/a_*)$} \\}
\startdata
 -1  & $720^3$ & $1024$ & 0.0014 & 0.026, 0.11, 0.19, 0.21, 0.24, 0.30 \\
-2 & $32^3$  & $64$ & 0.028 & 0.052, 0.13, 0.23, 0.31, 0.42, 0.49\\
   & $256^3$ & $512$ & 0.010 & 0.031, 0.054, 0.17, 0.29, 0.51, 0.67\\
   & $1024^3$& $2048$ & 0.006 & 0.024, 0.043, 0.063, 0.17, 0.30, 0.36 \\
-2.25 & $1584^3$ & $2048$ & 0.009 & 0.018, 0.021, 0.039, 0.084, 0.221,0.394\\
-2.5 & $720^3$ & $1024$ & 0.015 & 0.033, 0.074, 0.15, 0.23, 0.26, 0.29, 0.33 
\enddata
\label{tab:sims}
\end{deluxetable}

The ratio of the dimensionless power spectrum at the Nyquist
frequency, $k_{Ny}$, to that at the box scale, $k_B$, provides a
measure of a simulation's dynamic range.  For a scale-free power
spectrum
\begin{equation}
\frac{\Delta^2\left (k_{Ny}\right )}{\Delta^2\left (k_B\right )}
= \left (\frac{k_{Ny}}{k_B}\right )^{n+3} = N_m^{n+3}~.
\end{equation}
Thus, if $N=256^3$ particles are required to achieve a scaling
solution over a reasonable range in $\Delta^2$ when $n=-2$,
\citep{jain98}, $256^4\simeq1625^3$ particles are required at
$n=-2.25$, $256^6$ particles are required for $n=-2.5$, and $256^{12}$
particles are required for $n=-2.75$.  We set the particle number for
the $n=-2.25$ simulation on the basis of these arguments.

We fully anticipated that self-similar scaling would not be achieved
in our $n=-2.5$ simulation.  Our run, carried out with $N=720^3$
illustrates the difficulties that arise as one
attempts to simulate highly negative spectral indices.  In practice
$N=1584^3$ is the largest simulation we can perform within
word-addressing limits.  It is also clear from this discussion that
running an $n=-2.5$ simulation at $1584^3$ will yield little
improvement over our $720^3$ simulation since, apparently, one
requires $N=65536^3$.  Further discussions of the difficulties in
simulating scale-free models with $n\to -3$ can be found in
\citet{smith} and \citet{elahi}.

\section{Results}

\begin{figure}
\epsscale{1.0}
\plotone{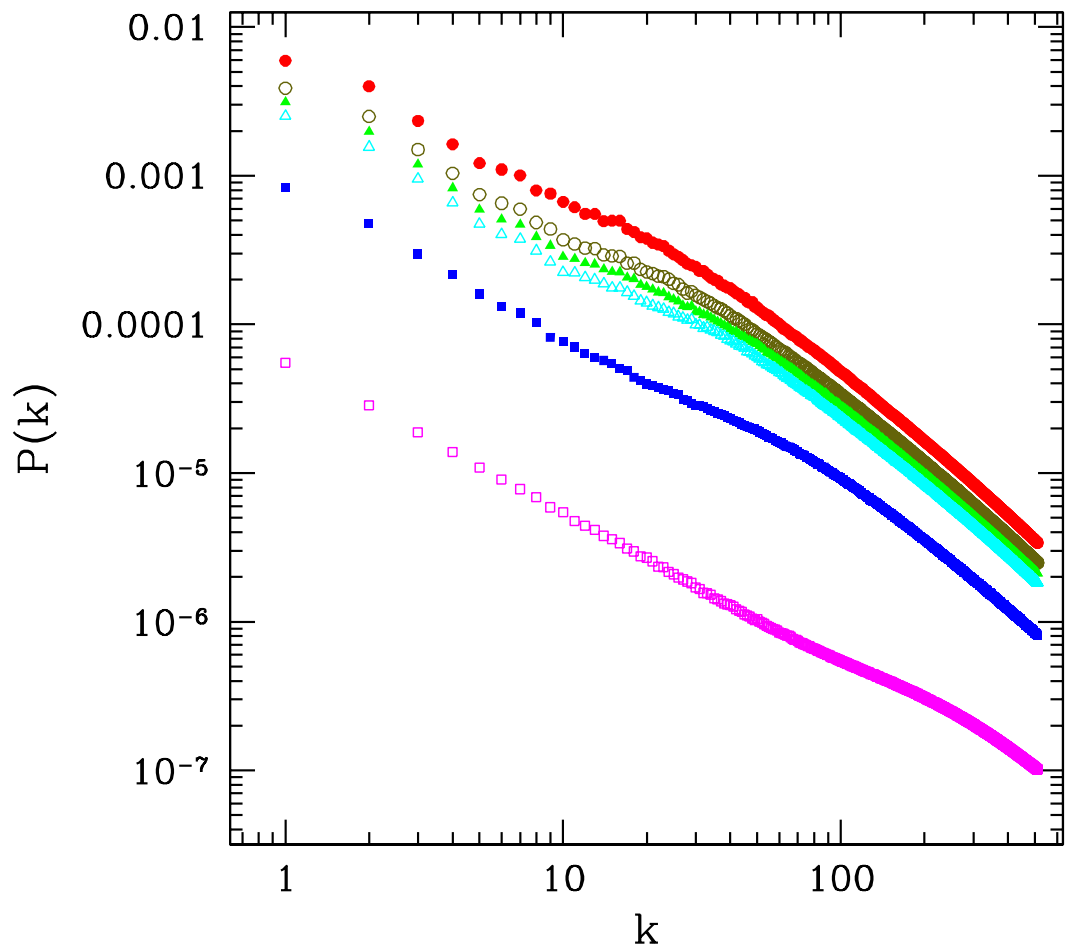}
\caption{Power spectrum, $P(k)$ as a function of wavenumber $k$ for
  the $n=-1$ simulation.  Different colours and symbols
  correspond to different output times as given in Table
  \ref{tab:sims}.  The sequence, from early to late times is
  magenta-blue-cyan-green-brown-red, or, alternatively, open
  square-filled square-open triangle-filled triangle-open
  circle-filled circle.}
\label{fig:pkn1}
\end{figure}

\begin{figure}
\epsscale{1.0}
\plotone{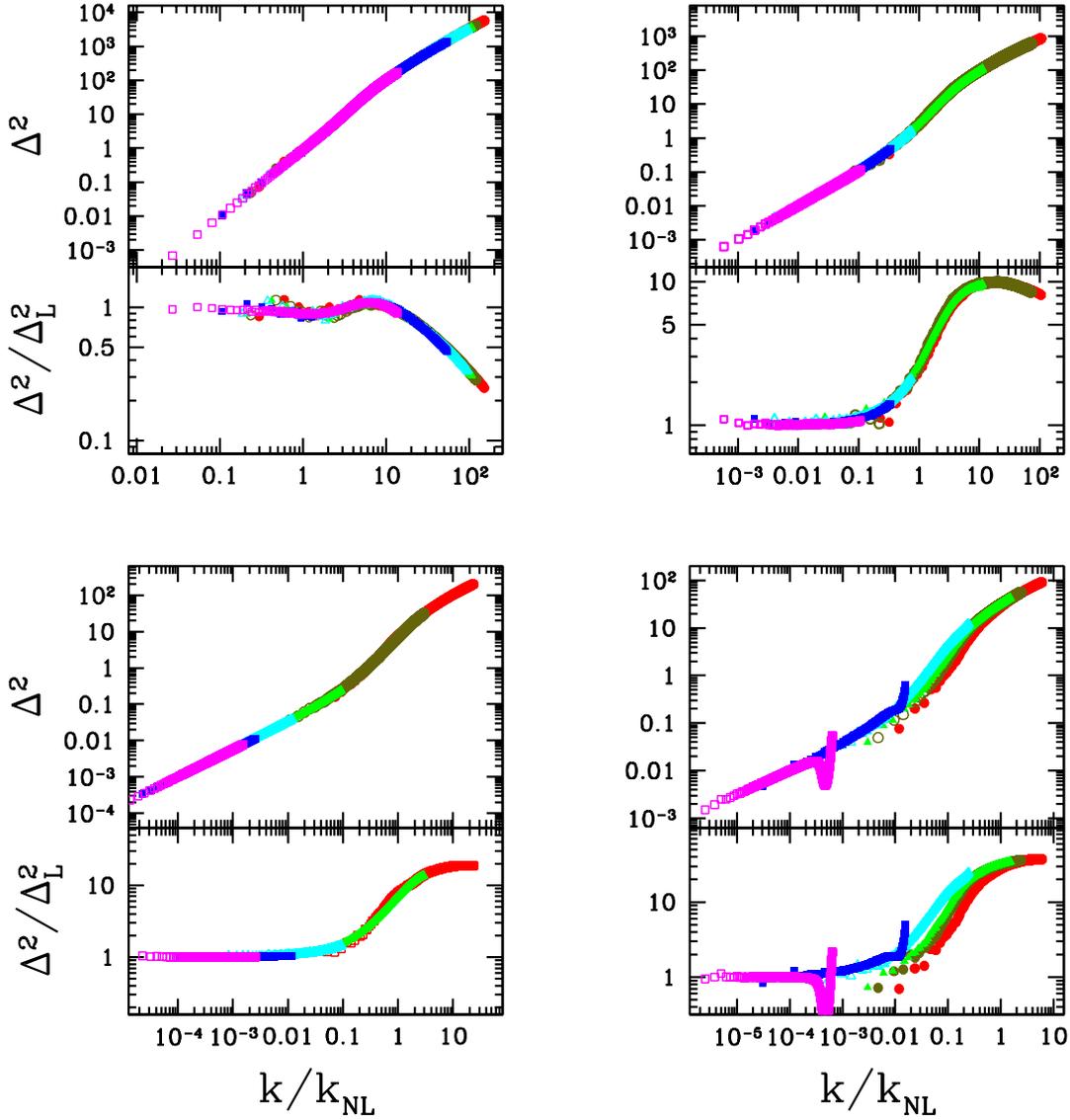}
\caption{Dimensionless power spectrum, $\Delta^2$ as a function of
  wavenumber $k$.  Colours and symbols are the same as in Figure
  \ref{fig:pkn1}.  Bottom panel in each quadrant shows the ratio
  $\Delta^2/\Delta_L^2$.  Upper left --- $n=-1$; Upper right --- $n=-2$;
  Lower left --- $n=-2.25$; Lower right --- $n=-2.5$.}
\label{fig:del}
\end{figure}

In Figure \ref{fig:pkn1} we plot the power spectrum from our $N=720^3$
$n=-1$ simulation at the six epochs listed in Table 1.  The power at a
given wavenumber increases with time while $k_{NL}$ (in this figure,
the wavenumber where the power spectrum deviates from the linear form,
$P(k)\propto k^{-1}$), decreases. These results are in close agreement
with previous simulations.  The cumulative halo distribution is also
in good agreement with the expected results \citep{elahi}.  In Figure
\ref{fig:del} we test the self-similar scaling ansatz by plotting the
dimensionless power spectra as a function of $k/k_{NL}$ for each of
our four high-resolution simulations.  Taken together, the spectra
from different epochs yield a composite power spectrum.  Consider,
first, the case $n=-1$ (upper left panel).  The power spectrum covers 7
orders of magnitude in $\Delta^2$ or, equivalently, 3-4 orders of
magnitude in $k$.  The fact that the composite power spectrum is very
nearly a single-valued function of $k/k_{NL}$ indicates that
self-similar scaling is essentially achieved.  Note also that
$\Delta^2/\Delta_L^2<1$ for most values of $k$.  This result is
consistent with PT (See Eq.\,\ref{eq:deltaPT} and note that
$\lambda(n=-1)<0$).

As with the $n=-1$ run, the composite spectra for $n=-2$ and $n=-2.25$
show excellent consistency with the scaling hypothesis.  However, a
departure from self-similar scaling is observed at large $a$ in the
$n=-2.5$ simulation highlighting the difficulty of simulating
$n\rightarrow -3$ spectral indices.  Note that the high-$k$ feature in
some of the early timesteps of our $n=-2.5$ simulation is a remnant of
the grid used in setting up the initial conditions.  The feature is
subdominant to the physical small-scale power at later times in the
$n=-2.5$ simulation.

\begin{figure}
\epsscale{1.0}
\plotone{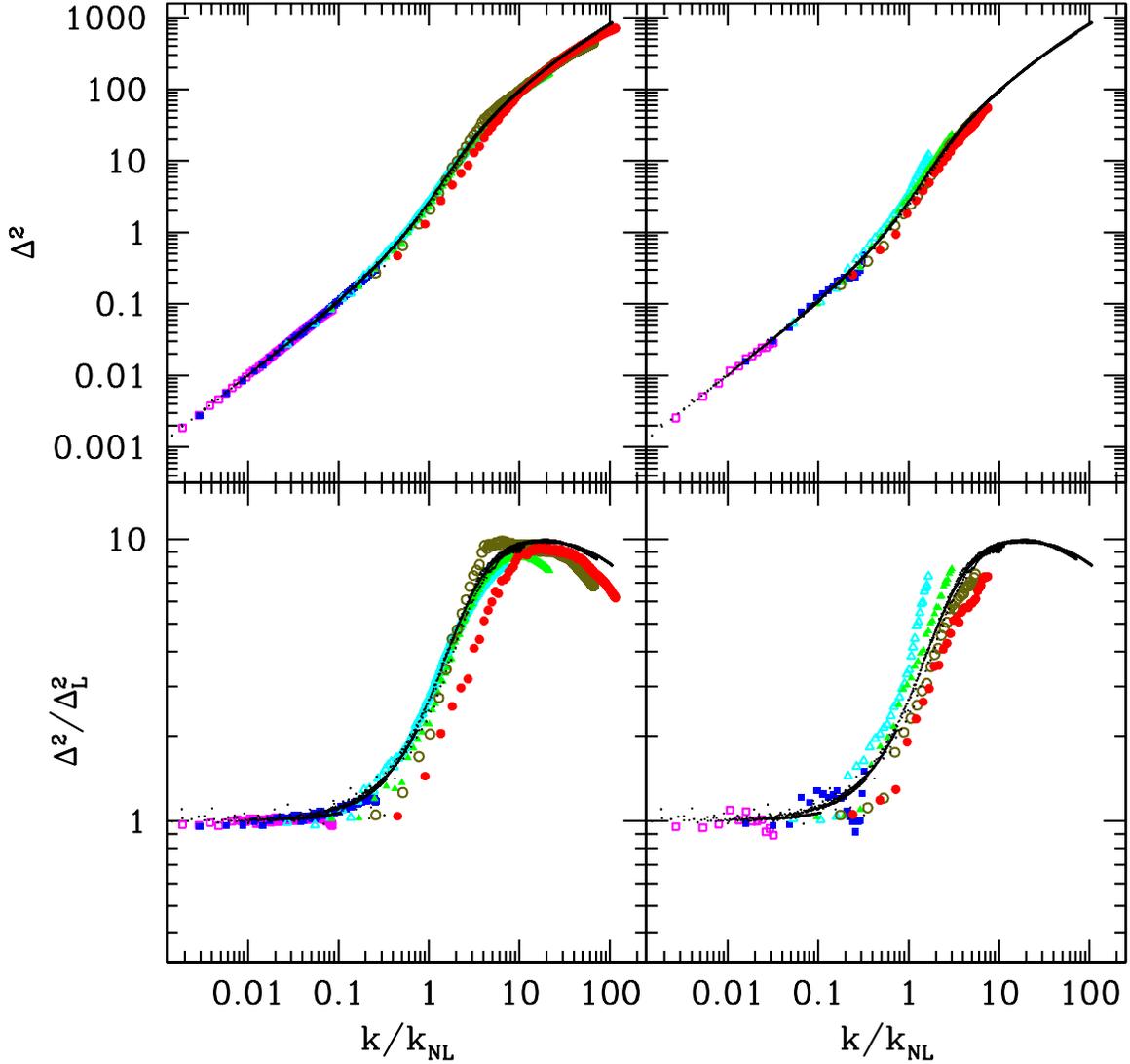}
\caption{Dimensionless power spectrum, $\Delta^2$, and the ratio
  $\Delta^2/\Delta^2_L$ for $n=-2$ from simulations with different
  numbers of particles.  Black points are power spectra at different
  timesteps measured in our highest resolution $(N = 1024^3)$
  simulation.  Superimposed in colour are measurements from the
  $N=256^3$ (left) and $N=32^3$ (right) simulations.  As in Figure
  \ref{fig:pkn1}, the sequence, from early to late times is
  magenta-blue-cyan-green-brown-red or, alternatively, open
  square-filled square, open triangle-filled triangle-open
  circle-filled circle.}
\label{fig:deln2lowres}
\end{figure}

The effect of resolution in achieving self-similar scaling is
illustrated in Figure \ref{fig:deln2lowres} where we compare the
spectra from the three $n=-2$ simulations.  A departure from
self-similar scaling is apparent in the $N=256^3$ simulation and quite
severe for $N=32^3$.  It may be that these simulations are
  over-evolved, as suggested by Figure \ref{fig:sigmamiss}.  In any case,
based upon the scaling arguments presented in the previous section, an
$N=32^3,~n=-2$ simulation should be comparable to an $N\simeq
32^6=1024^3,~n=-2.5$ simulation. Hence, it is not surprising that the
departures from self-similar scaling seen in the right-hand panels of
Figure \ref{fig:deln2lowres} are comparable to those seen in our
$n=-2.5$ simulation (lower-right panel of Figure \ref{fig:del}.  The
departure manifests itself in a suppression of power at small $k$ or
large scales, as expected since power is missing due to the finite
size of the simulation volume.

\subsection{Perturbative or Quasilinear Regime}\label{qlregime}

We now focus on the quasilinear regime in order to illustrate the
improvement renormalization group methods brings to perturbation
theory.  We implement the RG approach by solving the
  Eq.\,\ref{eq:rgequation} assuming an initial scale-free power
  spectrum with $n=-1,\,-2,\,-2.25$ or $-2.5$.  The initial spectrum is
  ``evolved'' forward in time (or equivalently, scale factor $a$)
  using a 4th-order Runge Kutta scheme with an adaptive stepsize (see,
  for example, \citet{press}).  Each Runge Kutta step requires an
  evaluation of the $\tilde P_{13}$ and $\tilde P_{22}$ integrals.
  These integrals have the same functional form as those that appear
  in \citet{sco96} except that $P_L$ is replaced by $\tilde P$ which,
  in turn, is updated at each step.  As with N-body simulations, we
  must truncate the initial power spectrum at both high and low
  wavenumbers.  Otherwise, the integrals would diverge.  \citet{sco96}
  use sharp cutoffs which are convenient for power-law spectra with
  integer $n$ since analytic expressions can be derived.  Smooth
  cutoffs are more manageable for the RG analysis where
  $\tilde{P}_{13}$ and $\tilde{P}_{22}$ must be evaluated numerically
  (Scoccimarro, {\it private communication}).

To make contact with our discussion in Section 2.2 we refer to the IR
and UV cutoffs as $k_B$ and $k_c$, respectively and assume an initial
power spectrum of the form
\begin{equation}
P_L(k) = Aa^2 k^n f_{UV}\left (k,k_c,\Delta \right)
f_{IR}\left (k,k_B,\Delta \right)
\end{equation}
where
\begin{equation}
  f_{IR}\left (k,\,k_B,\,\Delta k 
\right)=
  \frac{1}{2}
  \left (
    {\rm erf}\left (\frac{\ln{\left (k/k_B\right )}}
      {\Delta k}\right )+1\right )
\;\;\;\;\;
f_{UV}\left (k,\,k_c,\,\Delta k \right)=
 \frac{1}{2}\,
  {\rm erfc}\left (\frac{\ln{\left (k/k_c\right )}}{\Delta 
k}\right )~.
\end{equation}
The ratio $k_c/k_B$, which corresponds to the dynamic range of the
calculation, is set to $10^6$ while $\Delta$, which determines the
sharpness of the k-space cutoffs, is set equal to $0.5$.  For $n<-1$,
$P_{13}$ and $P_{22}$ have terms of order $\left (k_B/k_c\right
)^{n+1}$ which cancel, leaving behind a residual term of order
$k^{2n+3}$.  The challenge, numerically, is to determine the surviving
terms which can be much smaller than the terms that cancel, especially
for large $k$ and small $n$.  Here, we use the Romberg integration
routine from \citet{press}.

The solution to Eq.\,\ref{eq:rgequation} yields an evolutionary
sequence for the power spectrum, $\tilde P(k,\,a)$.  Departures from
the linear power spectrum increase with $a$ beginning at high
wavenumber.  We evolve $\tilde P$ until the $k_{NL}$ is roughly equal
to the geometric mean of $k_c$ and $k_B$.  Since our dynamic range is
a full three orders of magnitude greater than is found in our N-body
simulations, finite box effects are much less a concern here.  Moreover,
our results are insensitive to the form of the cutoff functions,
$f_{UV}$ and $f_{IR}$, since they are derived in a region well inside
the computation box.

\begin{figure}
\epsscale{1.0}
\plotone{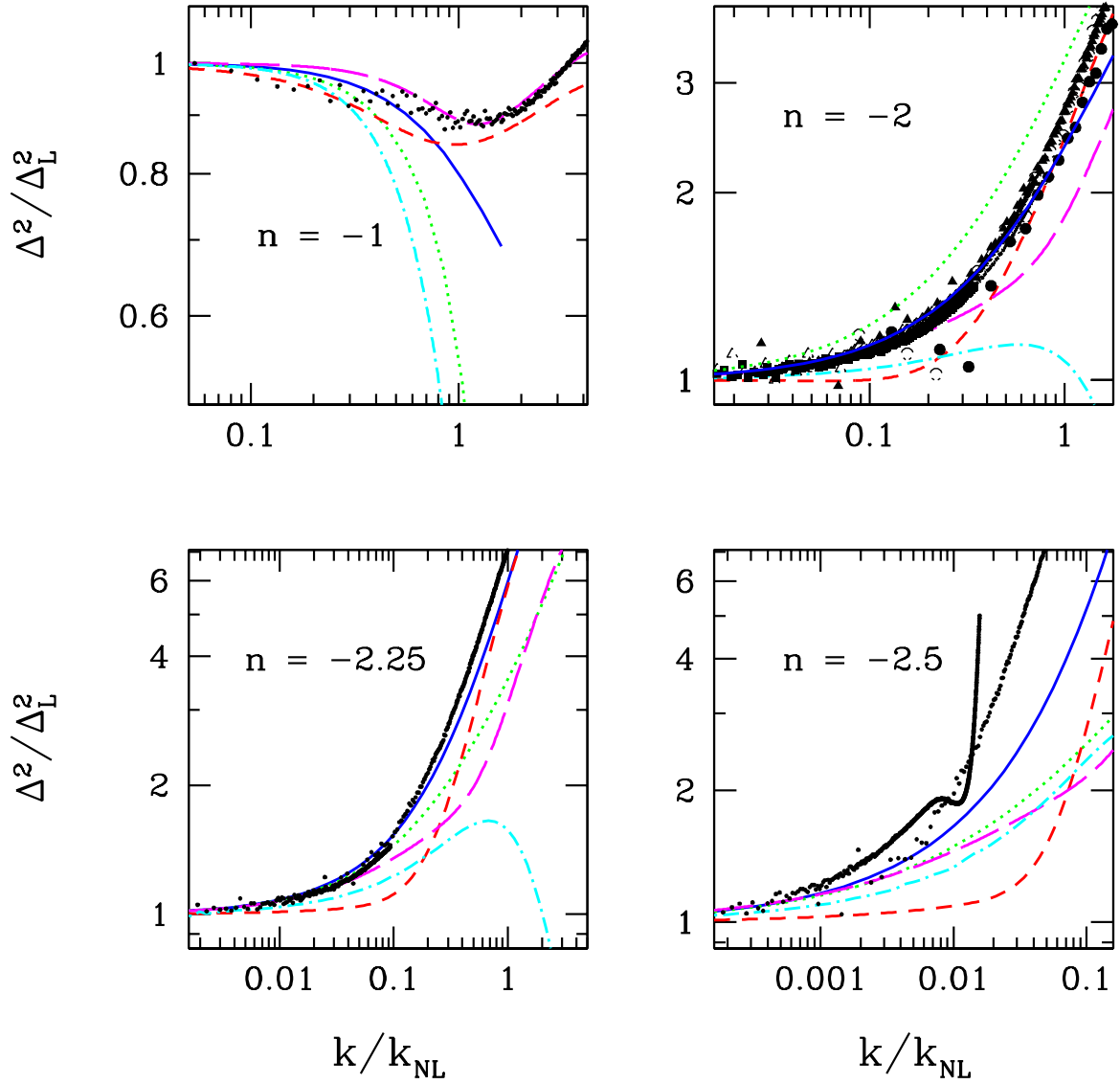}
\caption{The ratio $\Delta^2/\Delta_{L}$ as a function of $k/k_{NL}$
  from our four high-resolution simulations as labelled in each panel.
  Black points are from the simulation with the different symbols
  representing results from different outputs (from early to late
  outputs: solid squares, open triangles, solid triangles, open
  circles, solid circles). Line colours/types are: blue/solid --- RG;
  green/dotted --- one-loop; cyan/dot-dashed --- Zel'dovich
  approximation; red/short-dashed --- \citet{smith} fitting formula;
  magenta/long-dashed --- \citet{pd96} fitting formula.}
\label{fig:rgzoom}
\end{figure}

In Figure \ref{fig:rgzoom} we show the measured $\Delta^2$ in the
mildly nonlinear regime together with predictions from PT, RG-improved
PT, and the Zel'dovich approximation.  The latter is discussed in
\citet{taylor}.  Also shown are the fitting formulae of \citet{pd96}
and \citet{smith}.  Note that in the $n=-1$ case,
$\Delta^2/\Delta_{NL}^2$ slowly decreases with increasing $k$ for
$k\la k_{NL}$ and it is difficult to discern the true self-similar
evolution of the power spectrum given that the actual initial power
spectrum has a large-$k$ cutoff.  On the other hand, for $n=-2,\,$ and
$-2.25$, it is clear that RG does the best job of tracking the power
spectrum in the quasilinear regime the RG power spectra does not quite
capture the rapid evolution of the measured power spectra.  The
situation is less clear for $n=-2.5$ where the validity of the
simulation is in doubt.  The question remains as to whether agreement
between RG and the simulations might be improved by refinements in the
RG analysis, such as those suggested by \citet{mcdonald}.

\subsection{Nonlinear Regime}

\begin{figure}
\epsscale{1.0}
\plotone{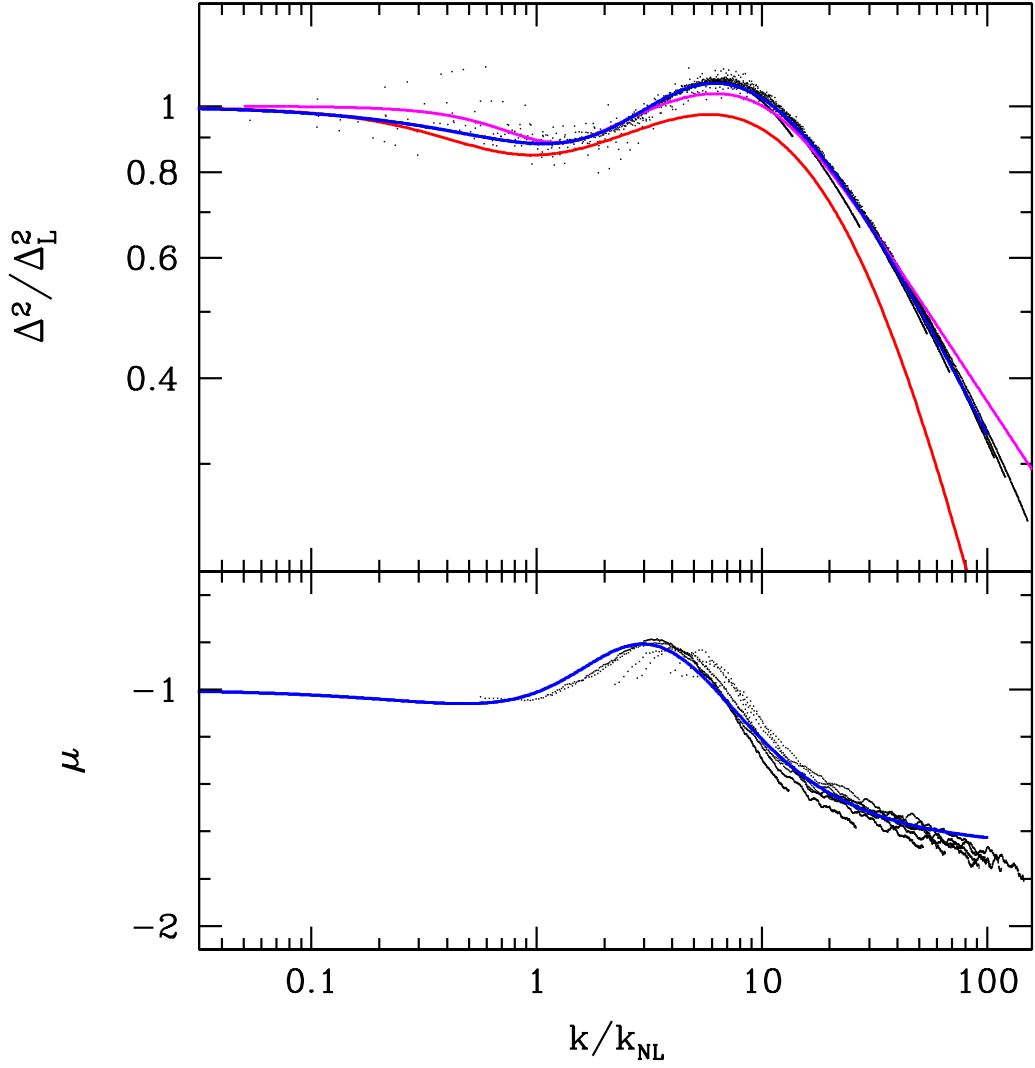}
\caption{The ratio $\Delta^2/\Delta_{L}$ as a function of $k/k_{NL}$
  for the $n=-1$, $N_m=1024$ simulation.  Shown is the full range in
  $k$ probed by the simulation is shown.  Black points are from the
  simulation.  Red curve is the \citet{smith} fitting formula.
  Magenta curve is the Peacock and Dodds fitting formula.  Blue curve
  is our own fitting formula, Eq.\,\ref{eq:ourfit}.  Plotted in the lower
  panel is the logarithmic slope, $\mu$ of the power spectrum (see
  text).}
\label{fig:delration1}
\end{figure}

In Figure \ref{fig:delration1}, we plot the complete power spectrum
for our high-resolution, $n=-1$ simulation together with the
predictions of \citet{pd96} and \citet{smith}.  Also shown is our own
fitting formula given by
\begin{equation}\label{eq:ourfit}
  \Delta^2(k)  = 
\Delta_L^2(k) g(k/k_{\rm NL})
\end{equation}
where
\begin{equation}
g(x) = \left (\frac{1 + Ax + Bx^\alpha}{1 + Cx^\gamma }\right )^\beta~.
\end{equation}
The form of this formula is motivated by that of \citet{pd96} but has
one additional parameter to allow for a more general behaviour in the
$k\to \infty$ limit.  The parameters, derived by performing a
nonlinear least-squares fit (see, for example, \citet{press}), are
given in Table \ref{tab:params}.  The lower panel in Figure
\ref{fig:delration1} shows the logarithmic slope $\mu \equiv d\ln
P/d\ln{k} = d\ln{\Delta^2}/d\ln{k} - 3$.  Note that $\mu$
monotonically decreases with $k$ near the Nyquist wavenumber.

\begin{figure}
\epsscale{1.0}
\plotone{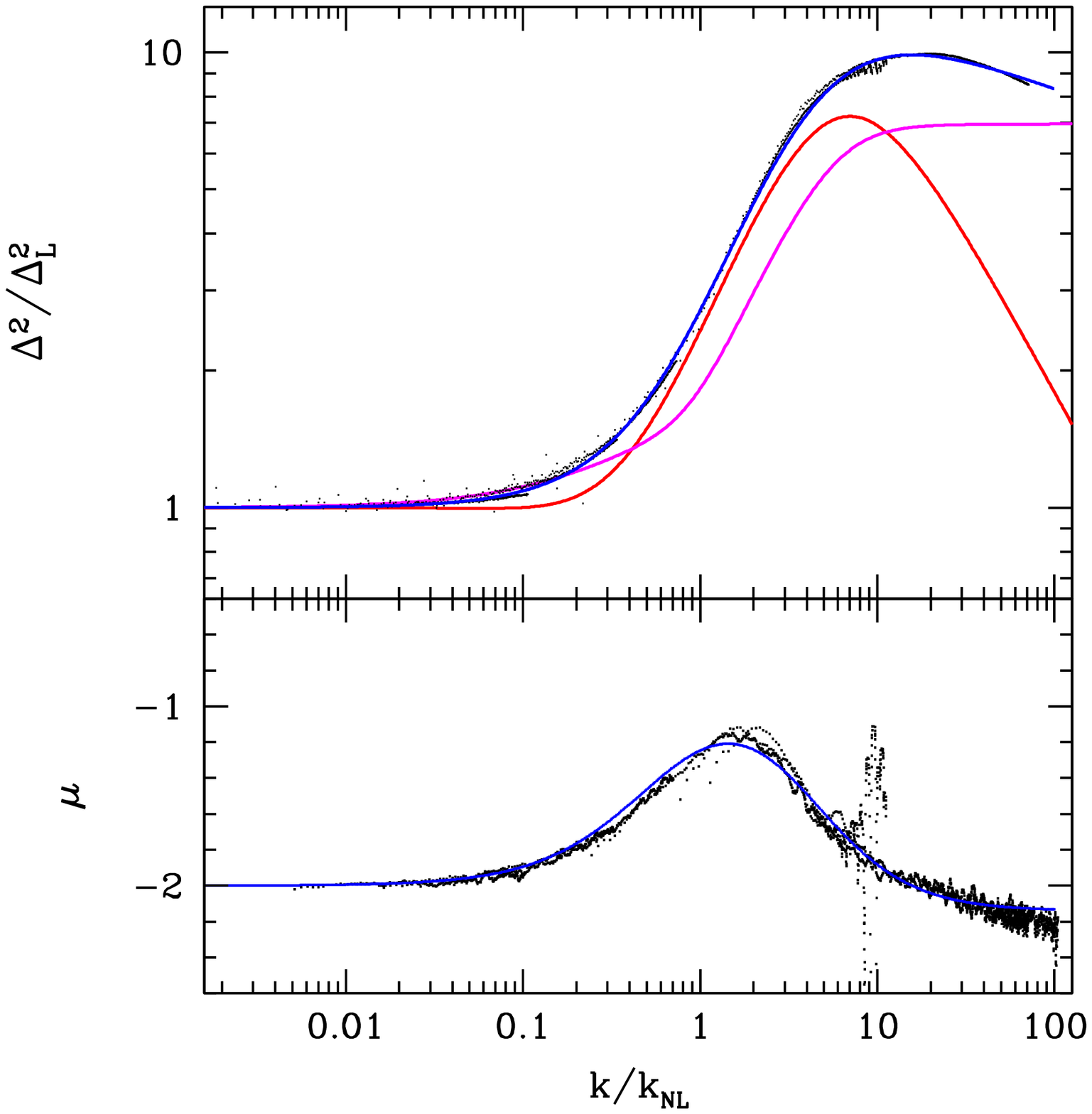}
\caption{Same as Figure \ref{fig:delration1} but for $n=-2$.}
\label{fig:delration2}
\end{figure}
\begin{figure}
\epsscale{1.0}
\plotone{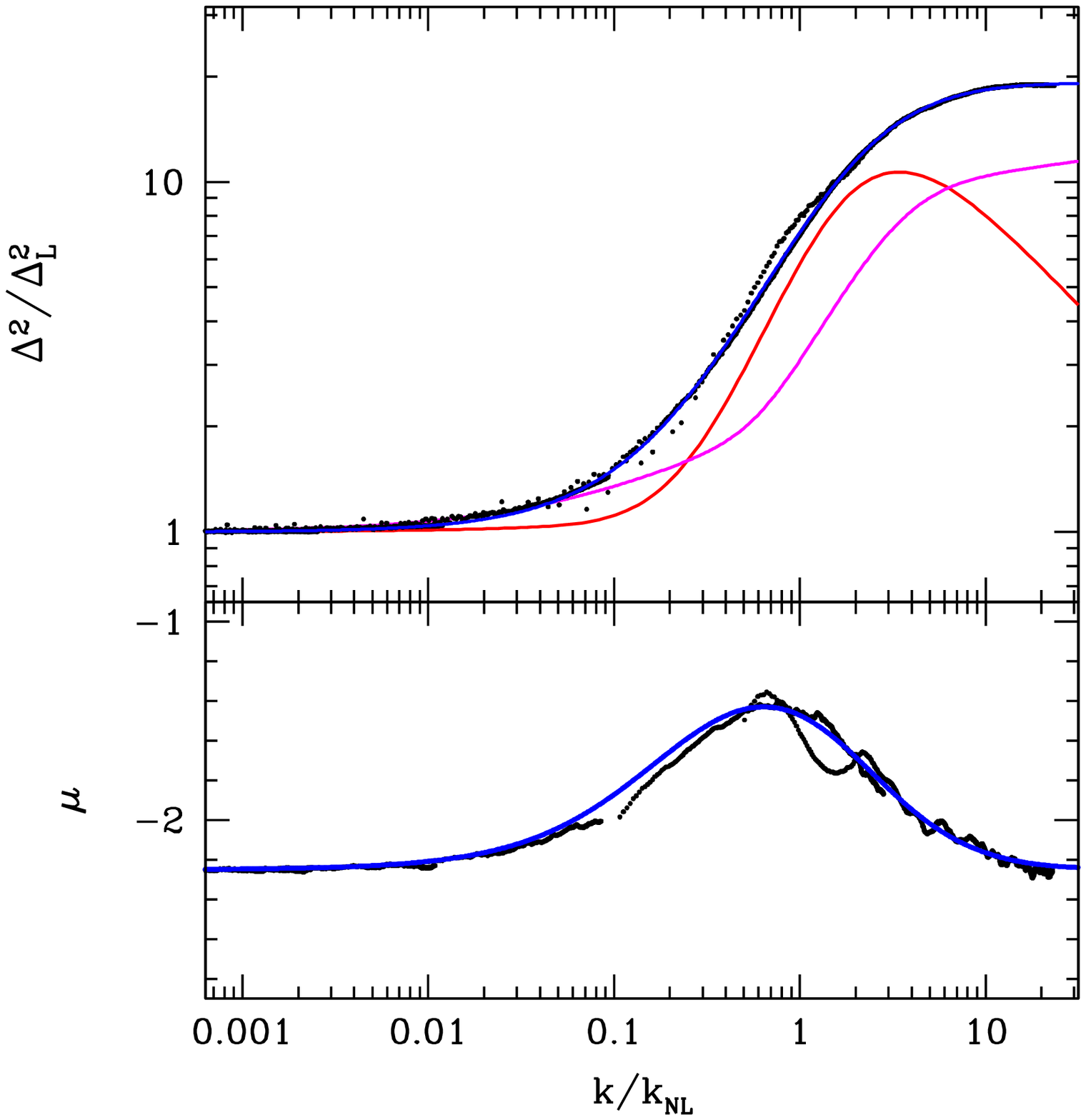}
\caption{Same as Figure \ref{fig:delration1} but for $n=-2.25$.}
\label{fig:delration225}
\end{figure}

In Figures \ref{fig:delration2} and \ref{fig:delration225}, we show
$\Delta^2$ and $\mu$ for our high-resolution, $n=-2$ and $-2.25$
simulations.  Again, $\mu$ decreases monotonically in the large-$k$
limit.  Nevertheless, by design, virtually all fitting formulae
(including our own) have a power-law form in the high-$k$ limit, that
is, $P(k)\propto k^{{\bar\mu}}$ as $k\to \infty$.  The lesson is that
fitting formula should not be extrapolated to scales below the
smallest scales probed by the simulation used in their construction.

Neither the \citet{pd96} nor \citet{smith} fitting formulae do a
particularly good job of fitting the power spectra from our
simulations.  For $n=-2$, the \citet{smith} formula provides a
reasonable fit up to $k/k_{NL}\simeq 5$ but decreases too rapidly
beyond this point.  Conversely, \citet{pd96} predict that
$\Delta^2/\Delta_L^2$ is constant in the large-$k$ limit whereas the
measured power spectrum shows a clear decline.  The discrepancies
between predicted and measured power spectra for $n=-2.25$ are equally
severe.  By contrast, Eq.\,\ref{eq:ourfit} provides an excellent fit
to the nonlinear power spectra from our high-resolution simulations.
And while it has one more parameter than the fitting formula of
\citet{pd96}, it has two fewer than that of \citet{smith}.

\begin{figure}
\epsscale{1.0}
\plotone{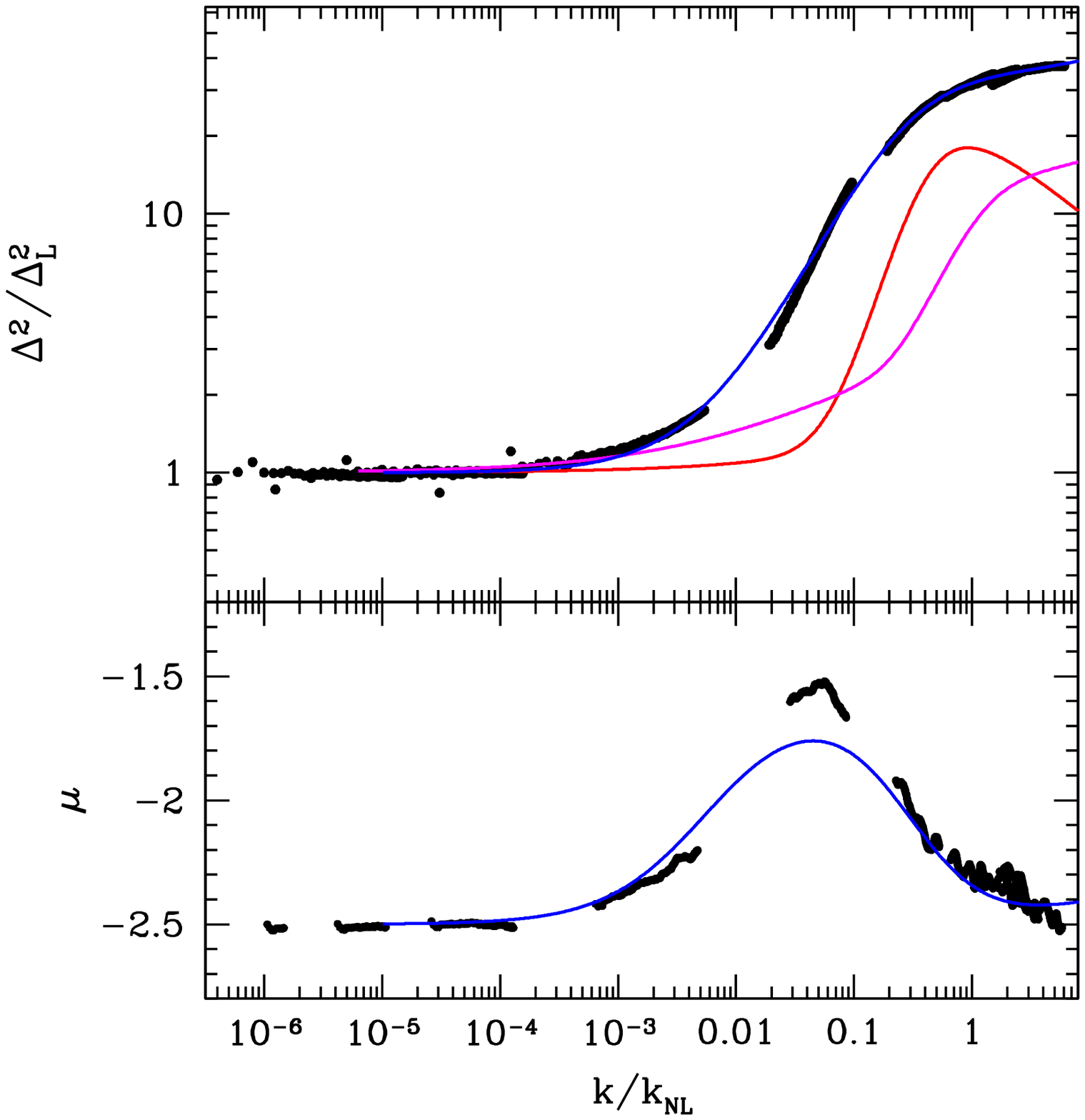}
\caption{Same as Figure \ref{fig:delration1} but for $n=-2.5$.}
\label{fig:delration25}
\end{figure}

The $n=-2.5$ case, shown in Figure \ref{fig:delration25}, is difficult
to analyse because of the departure from self-similar scaling.  In
this plot, we truncate the power spectra from different timesteps at
large $k$, where the effects of aliasing is apparent, and at small
$k$, where the effects of missing large-scale power is apparent.  We
contend that this procedure yields a roughly continuous curve, which
provides a reasonable facsimile of the true power spectrum.  The
plausibility of this procedure is illustrated in the left-hand panels
of Figure \ref{fig:deln2lowres} where one can imagine carrying out a
similar procedure with our $n=-2$, $N=256^3$ results to yield an
approximate form for the power spectrum from our high-resolution
simulation.

\begin{deluxetable}{ccccccc}
\tablecaption{Parameters for fitting formula, Eq.\,\ref{eq:ourfit}}
\tablewidth{0pt}
\tablehead{
\multicolumn{1}{c}{$n$} & \multicolumn{1}{c}{$A$} &
\multicolumn{1}{c}{$B$} & \multicolumn{1}{c}{$C$} &
\multicolumn{1}{c}{$\alpha$} & \multicolumn{1}{c}{$\gamma$} &
\multicolumn{1}{c}{$\beta$} 
}
\startdata
-1 & -0.158 & 0.181 & 0.0729 & 1.571 & 1.845 & 2.664 \\
-2 & -0.0312 & 0.690 & 0.478 & 1.243 & 1.266 & 8.647 \\
-2.25 & 3.471 & 4.038 & 0.348 & 1.413 & 1.372 & 0.659 \\
-2.5 & 174.0 & 110.2 & 4.532 & 1.492 & 1.231 & 0.879 
\enddata
\label{tab:params}
\end{deluxetable}

\section{Halo Model Revisited}

In this section, we explore the halo model vis-\`{a}-vis our
simulation results in more detail.  Our focus here is on the 
high-$k$ limit of the nonlinear power spectrum where the one-halo
term dominates.
The term can be expressed as an integral over the halo mass
function, $dn/dM$, and the Fourier-transformed density profile of a
single halo, $\tilde\rho\left (k,\,M\right )$.  Following
\citet{seljak}, we use the peak height $\nu\equiv \left (
  \delta_c/\sigma\left (M\right )\right )^2$ as the integration
variable where $\delta_c$ is the critical overdensity for spherical
collapse ($\delta_c\simeq 1.68$ in an Einstein-de Sitter universe) and
$\sigma(M)$ is the rms mass overdensity for a spherical region of
radius $R = \left (3M/4\pi\rho_{\rm bg}\right )^{1/3}$.  For
scale-free models, $M\propto \nu^{3/\left (n+3\right )}$.
One finds
\begin{equation}
P_{1h}(k) = \frac{1}{\left (2\pi\right )^3}
\int h\left (\nu\right )\left (\frac{M}{\rho_{\rm bg}}\right )
\left [\frac{\tilde{\rho}\left (k,\,M\right )}{M}\right ]^2\,d\nu~.
\end{equation}
where
\begin{equation}
h\left (\nu\right ) = \frac{M}{\rho_{\rm bg}}\frac{dn}{dM}\frac{dM}{d\nu},
\end{equation}
is a dimensionless form for the halo mass function.

A commonly used fitting formulae for halo density profiles take the
form
\begin{equation}\label{eq:densityprofile}
\rho(r) = \frac{\rho_0}{\left (r/r_s\right )^\gamma
\left ( 1 + r/r_s\right )^{\eta-\gamma}},
\end{equation}
where $\eta\simeq 3$ and $\gamma\simeq 0.5-1.5$ \citep{nfw,moore99b,
  kravtsov}.  The characteristic halo scale length, $r_s$, depends on
the halo mass through the relation $r_s = r_{\rm vir}/c_{\rm vir}$
where $r_{\rm vir}$ is the virial radius and $c_{\rm vir}$ is the halo
concentration parameter.  Cosmological simulations indicate that the
concentration parameter decreases with mass, roughly as a power-law
(see, for example, \citet{nfw} and \citet{bullock01}).  With this in
mind, we follow \citet{seljak} and \citet{mafry} and write $c_{\rm
  vir} = \left (M_{\rm vir}/M_0\right )^{-\beta}$ so that $r_s\propto
M^{\left (1+3\beta\right )/3}$.

Our focus is on the power spectrum in the high-$k$ limit where the
Fourier transform of $\rho(r)$ may be approximated by a step function,
\begin{equation}
\tilde{\rho}(k,\,M)\simeq M\Theta\left (1-kr_s(M)\right )~.
\end{equation}
(At wavenumbers above $k=r_s^{-1}$, $\tilde\rho$ decreases as $\left
  (kr_s\right )^{3-\gamma}$ but the Heaviside function provides a
suitable form for our discussion.)  \citet{pressschechter} and
\citet{shethtormen} provide analytic expressions for $h(\nu)$.  In the
small-$\nu$ limit (i.e., small $M$ or large $k$ limit), one finds $\nu
h\propto\nu^{\alpha}$ where $\alpha=0.5$ ($\alpha=0.3$) for
\citet{pressschechter} (\citet{shethtormen}).
Putting all this together, we find that in the large-$k$ limit
$\Delta^2_{1h}\propto k^{\bar{\mu}_{1h}+3}$
where 
\begin{equation}\label{asymptotic}
\bar{\mu}_{1h}=\frac{9\beta - \alpha\left (n+3\right )}{1+3\beta}-3
\end{equation}
\citet{mafry}.  Note, however, that ${\bar{\mu}}_{1h}$ is independent
of the cusp parameter $\gamma$ indicating that the power spectrum in
the strongly nonlinear regime is insensitive to the structure of the
inner halo.

\begin{figure}
\epsscale{1.0}
\plotone{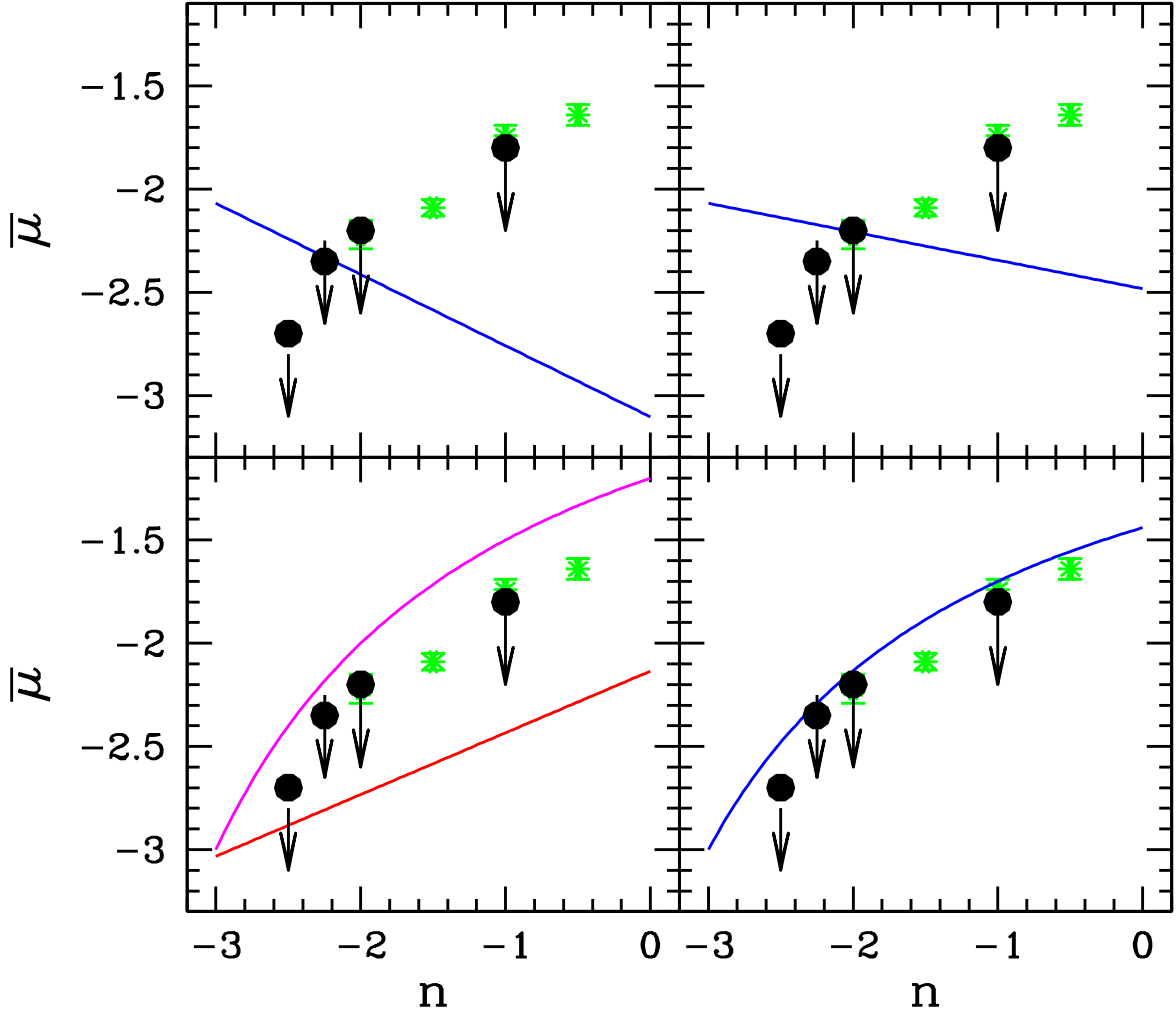}
\caption{Asymptotic slope of the power spectrum, $\bar{\mu}$, as a
  function of the slope, $n$, of the linear power spectrum.  The black
  points are from our simulations while the green points are from the
  \citet{smith} simulations.  In the upper left panel we show the
  predictions of the halo model (Eq.\,\ref{asymptotic}) for $\beta =
  0.15$ and $\alpha = 0.5$ (Press \& Schechter).  The upper right
  panel shows the prediction of the halo model assuming $\beta=0.15$
  and $\alpha = 0.2$ (Sheth \& Tormen).  The lower left panel shows
  the predictions from \citet{pd96} (magenta curve) and \citet{smith}
  (red curve).  The lower right panel, shows the predictions of the
  halo model using the prescription from \citet{bullock01} for
  $\beta$.}
\label{fig:asymslope}
\end{figure}

In Figure \ref{fig:asymslope} we plot our results for the asymptotic
slope of the nonlinear power spectrum, $\bar{\mu}$, together with
those from the \citet{smith} simulations.  We make the point of
displaying our results as upper bounds on $\bar{\mu}$ since the slope
of $\mu$ appears to be a decreasing function of $k$ as $k\to k_{Ny}$
(see Figures \ref{fig:delration1}-\ref{fig:delration25}).  For $n=-1$
and $-2$, these bounds agree with the quoted values from the
\citet{smith} simulations.  Furthermore, the functional dependence of
$\bar{\mu}$ on $n$ from their fitting formula appears to be consistent
with our $n=-2.25$ and $n=-2.5$ results.

Our results suggest that $\mu$ increases with increasing $n$ and that
$\bar{\mu}\left (n\to -3\right )=-3$.  In other words, as $n\to -3$,
the nonlinear dimensionless power spectrum becomes independent of $k$
(i.e., equal power per logarithmic wavenumber bin) just as with the
linear power spectrum.  Formulations of the halo model with constant,
nonzero $\beta$ cannot reproduce this behaviour.  To illustrate this
point, we plot these predictions for $\bar{\mu}_{1h}$ assuming
$\beta=0.15$, as in \citet{seljak}, and either the
\citet{pressschechter} or \citet{shethtormen} values for $\alpha$.
These predictions are inconsistent with our simulation results and
those of \citet{smith}.

Clearly, the dependence of the concentration parameter on halo mass is
central to the development of the halo model.  \citet{bullock01}
devised a toy model to explain the concentration-mass relation seen in
simulations.  In the case of scale-free cosmologies their model
predicts $\beta = (n+3)/6$ which, when combined with
Eq.\,\ref{asymptotic}, would seem to yield the desired behaviour for
$\bar{\mu}_{hh}$ in the $n\to -3$ limit.  The lower panel in Figure
\ref{fig:asymslope} shows this prediction.  While it does better
than constant-$\beta$ versions of the halo model, the predicted
$\bar{\mu}$ tends to lie above the values obtained in the simulation.

\section{Summary and Conclusions}

The fundamental tenet of the hierarchical clustering scenario is that
small-scale objects form earlier than large-scale ones.  A corollary
of this statement is that individual structures can be identified with
a specific wavenumber range of the primordial power spectrum according
to their mass.  In CDM cosmologies, the logarithmic slope or spectral
index of the primordial power spectrum runs from 1 at large scales to
-3 at small scales.  Thus, as we increase the dynamic range in our
simulations and push to smaller and smaller scales, we probe
structures that form from density perturbations with a power spectrum
approaching $k^{-3}$.  However this limit represents a singular case
where the dimensionless power spectrum is independent of scale and
structures across a wide range in mass collapse nearly simultaneously.
The nature of structure formation changes and the computing
requirements for performing simulations increase dramatically.

This work and our companion paper, \citet{elahi}, provide inside into
the underlying physics of $\Lambda$CDM models by considering
scale-free cosmologies.  We focus here on the nonlinear power spectrum
and in \citet{elahi}, on the distribution of subhaloes.  The evolution
of the power spectrum in scale-free cosmologies is remarkably simple
--- the dimensionless power spectrum, when written as a function of
the ratio $k/k_{NL}$, is time-independent.  Obviously in simulations,
the finite computation volume breaks the scale-free nature of the
problem and leads to departures from the scaling solution.  The
dimensionless power spectrum provides a simple test of whether finite
volume effects have corrupted the simulation \citep{jain98}.

Our high-resolution $n=-1, -2$ and $-2.25$ simulations demonstrate the
scaling solution across the simulation volume while showing clear
differences with simulations performed at lower resolution.  Moreover,
our results differ markedly from the the fitting formula provided by
\citet{pd96} and \citet{smith}.  A plausible power spectrum for
$n=-2.5$ was constructed by stitching together outputs from different
timesteps.  Though it shows significant, and entirely expected
departures from the scaling solution, it represents our best estimate
of the power spectrum for models with $n$ this small.  We summarize
our results for our four high-resolution simulations by means of a
simple fitting formula.  Future work will fill in the gaps in $n$
(e.g., $n=-1.25,\,-1.5, -1.75,$ and $-2.75$) and enable use to develop
a model for power spectra of arbitrary $n$ and therefore arbitrary
shape.

The renormalization group improvements to perturbation theory
developed in \citet{mcdonald} represent a promising avenue for
studying scale-free models with $n<-2$ and likewise, the low-mass
limit of the CDM hierarchy.  Not surprisingly, the calculations become
more difficult as $n\to -3$.  Our analysis of the $n=-2$ and
$-2.25$ cases confirms McDonald's claim that RG does lead to an
improvement in the predictions of perturbation theory when compared to
simulations with the caveat that, as $n$ becomes more negative, the
RG-predicted power spectrum fails to capture the rapid rise of the
power spectrum seen in the simulations.  Our analysis for the $n=-2.5$
case is less conclusive but departures in the simulations from the
scaling solution suggest that the problem may reside there rather than
in the PT analysis.  In principle, RG-improved PT can yield a handle
on the form of the power spectrum in the mildly nonlinear regime even
as $n\to -3$.

The halo model provides a theoretical framework for understanding the
two-point correlation function and nonlinear power spectrum.  Our
results, with respect to this model, are somewhat inconclusive.  We
agree with \citet{smith} that the stable clustering hypothesis of
\citet{hklm} and \citet{pd96} fails.  On the other hand, the halo
model appears to have difficulty reproducing the relation between the
asymptotic slope of the power spectrum and $n$.  We must therefore
settle for an empirical fitting formula for the nonlinear power
spectrum.

\acknowledgements{It is a pleasure to thank P. McDonald,
  R. Scocciamarro, and C. Orban for useful conversations.  We thank R.
  Smith for carefully reading our manuscript and providing valuable
  suggestions.  We also thank C. Orban for uncovering out an important
  typo in one of the equations.  PJE acknowledges financial support
  from the Natural Science and Engineering Research Council of Canada
  (NSERC). RJT and LW acknowledge funding by respective Discovery
  Grants from NSERC. RJT is also supported by grants from the Canada
  Foundation for Innovation and the Canada Research Chairs
  Program. Simulations and analysis were performed on the computing
  facilities at the High Performance Computing Virtual Laboratory at
  Queen's University, SHARCNET, Arizona State University Fulton High
  Performance Computing Initiative and the {\em Computational
    Astrophysics Laboratory} at Saint Mary's University.}

\end{document}